# Integrating Association Rules with Decision Trees in Object-Relational Databases

Maruthi Rohit Ayyagari

*College of Business, University of Dallas, Texas, USA*

E-mail: rayyagari@udallas.edu

*Abstract*—Research has provided evidence that associative classification produces more accurate results compared to other classification models. The Classification Based on Association (CBA) is one of the famous Associative Classification algorithms that generates accurate classifiers. However, current association classification algorithms reside external to databases, which reduces the flexibility of enterprise analytics systems. This paper implements the CBA in Oracle database using two variant models—hardcoding the CBA in Oracle Data Mining (ODM) package and Integrating Oracle Apriori model with the Oracle Decision tree model. We compared the proposed model performance with Naïve Bayes, Support Vector Machine, Random Forests, and Decision Tree over 18 datasets from UCI. Results showed that our models outperformed the original CBA model with 1% and is competitive to chosen classification models over benchmark datasets.

## I. INTRODUCTION

Data mining is a process of extracting useful information and hidden knowledge from large databases [1]. Data mining has four tasks; association, classification, clustering, and regression. In the last few years, classification and association have been used widely by data mining communities[2], [3] which result in a new approach in data mining known as the Associative Classification (AC).

The AC integrates associative rule discovery[4] and classification to predict a class label [5], [6]. Several studies have indicated that the AC algorithms can extract classifiers competitive with those produced by decision trees[7]–[9], rule induction[9], [10] and probabilistic approaches. Nowadays many algorithms are built based on the AC approach such as the Classification Based on Association (CBA) [11], [12], Classification Based on Multiple Class-Association Rules (CMAR) [13], Multi-Class Classification Based on Association Rules (MCAR) [14], multi-class, Multi-Label Associative Classification (MMAC) [15], and Classification based on Predictive Association Rules (CPAR) [16][17].

Data is currently being extended exponentially especially with the introduction of objects such images and composite attributes. Therefore new database models have emerged Object Databases is considered one data model that is being used in many enterprises.

Real data often includes noise including missing or incorrect values. Data should be filtered, normalized, sampled, transformed before model building (known as discretization). Some algorithms require data to be cleaned and preprocessed before mining. When the data is stored in a database, data discretization can be performed using database facilities efficiently. Therefore, analytic software becomes productive with online predictions. As a result, predictions become a lucrative market for large organizations.

Oracle database is an object-relational database management system (ORDBMS) that includes many features such as high scalability, high performance, and the availability on multiple platforms[18]. Oracle has produced an option implemented in the Oracle database kernel called the Oracle Data Mining (ODM) [19]. The ODM processes use built-in features of Oracle database to maximize scalability and make efficient use of the system resources. It contains the following data mining models: Apriori, Decision Tree, Generalized Linear Models, k-Means, Minimum Description Length, Naive Bayes, Non-Negative Matrix Factorization, O-Cluster and Support Vector Machines.

To the best of our knowledge, there is no AC algorithm such as CBA implemented in Oracle. When CBA is implemented in the ORDBMS it enables users to mine different datasets from the database directly; therefore, the CBA algorithm increases productivity. In this paper, we augment the power of Oracle Database: data availability, scalability, and performance with a promising data mining algorithm.

We propose to implement and integrate the CBA algorithm with ODM package. The new implementation is two folds. The first one is CBA implementation based on Oracle Apriori model and the second is based on integrating Oracle Apriori model with Oracle Decision Tree model.

The research aims to meet the following objectives:

1. Analyzing Oracle ODM package and find ways for possible integrations to CBA model.
2. Write needed source code that integrates with ODM Package.

3. Compare the accuracy of our models with Naïve Bayes, Decision Trees, SVM, and Random Forests models.

The following Section is a literature review about the AC and Decision trees. Section III illustrates CBA by an example. Section IV contains the proposed model whereas Section V has experimental results and its evaluation. Finally, we conclude the findings in Section VI.

## II. LITERATURE REVIEW

Classification can be described as a supervised learning algorithm in the machine learning process. It assigns class labels to data objects based on prior knowledge of the class where the data records belong.

Association rule mining and classification rule mining are fundamental techniques in data mining, and sometimes they are indispensable to practical applications. Therefore, integrating these two techniques is valuable and useful as integration will produce significant savings and conveniences to the users[12].

The associative classification (AC) approach was introduced by [5] to build classifiers (sets of rules) and later attracted many researchers, e.g. [12], [16], from data mining and machine learning communities. The AC is a case of association rule mining in which only the class attribute is considered in the rule's consequent (right-hand side of the if-then). For example, in a rule such as X➔Y, in the AC, the Y must be a class attribute. Empirical studies[3], [13] showed that AC often builds more accurate classification systems than traditional classification techniques. Moreover, unlike neural networks[20], which produce classification models that are hard to understand or interpret by end-users, the AC generates rules that are easy to understand and manipulate by end-users.

Liu has developed various variants of the CBA algorithm [12], [21], [22]. The standard CBA algorithm consists of two steps; candidate rules generation and classifier builder. In the candidate rules generation, a special subset of association rules, the class association rules (CARs), will be generated. The classifier will be built based on the discovered CARs. Liu has also implemented the CBA algorithm with multiple minimum support [6]. The improved CBA was able to use the most accurate rules for classifier building. Liu has applied a set of 34 benchmark datasets on his techniques. The results showed that the new techniques reduce the error of CBA by 17% and is superior to the previous CBA on 26 of the 34 datasets.

Positive and negative rules was introduced as an approach to building new classifiers [23]. The correlation coefficient which measures the strength of the linear relationship between two variables was used for pruning [23]. As he has used support and confidence for pruning in rules generation, he has used correlation coefficient pruning in classifier building. Results over six UCI datasets showed that negative association rules are useful when used with positive ones for producing competitive classification systems.

Many approaches have been adopted in the AC rule discovery[24]–[26], the FP-growth approach[27], and algorithms such as CPAR[16] that uses a greedy strategy presented in FOIL [10]. To conclude, the AC algorithms[3], [15] extends tid lists intersections methods of vertical association rule data layout[28] to solve classification benchmark problems.

The decision tree algorithm is a data mining induction techniques that recursively partitions a dataset of records using the depth-first greedy approach or breadth-first approach until all the data items belong to a particular class. At each node of the tree, a decision of best split is made using impurity measures [8]. The tree leaves are made up of the class labels which the data items have been a group. Decision tree classification technique is performed in two phases: the tree building and the tree pruning. The tree building is done in a top-down manner. During the first phase, the tree is recursively partitioned until all the data items belong to the same class label. Therefore, the first phase is very tasking and computationally intensive as the training dataset is traversed repeatedly. The second phase, the tree pruning is done a bottom-up fashion based on entropy or information gain.

## III. THE CBA ALGORITHM

Although CBA seems to be outdated, it is still used in many recent works [3] [29] [30]. The CBA is a classification algorithm based on association rules[12]. The CBA utilizes the association rules discovery algorithm, Apriori [31]. The Apriori generates association rules satisfying user-defined minimum support and minimum confidence thresholds.

An itemset is a set of transaction data that can be used in mining. The support of an itemset is defined as the number of transactions in the dataset that contain the item set [32]. Given an association rule X➔Y, confidence is defined as the total transactions containing Y given that it includes X. The CBA selects a subset of these association rules called class association rules (CARs). i.e., that is the target of the if-then rule is the class label. The CBA algorithm has three stages:

1. Generate frequent rule items CBA-RG
2. Apply the CBA-RG algorithm to generate the CARs set.
3. Build the classifier based on the CARs set.

The CBA algorithm is explained by an example [12]. Table I shows A, B are attributes, and C is the class label. Assume that the given minimum support is 15% and the minimum confidence is 60%, then follow results in Table II onwards.

TABLE I. DATASET SAMPLE

| A | B | C |
|---|---|---|
| e | P | y |
| e | P | y |
| e | Q | y |
| g | Q | y |
| g | Q | y |
| g | Q | n |
| g | W | n |
| g | W | n |
| e | P | n |
| f | Q | n |

Table II shows the frequent rule items denoted as $F_1$ and $F_2$ and candidate rule items denoted as $C_1$, $C_2$. The frequent rule items are rule items that satisfy minimum support. The rule item is represented in the algorithm in the form:

<(*condset*, *condsupCount*), (y, *rulesupCount*)>,

where *condset* is the condition (the if part), condsupCount is the support count, y is the class label, and *rulesupCount* the rule confidence.

Table III shows the class association rules, a rule that satisfies minimum support and confidence thresholds. Table IV shows the class association rules after pruning using database coverage heuristic that is used by the CBA. Table V shows the list of generated rules.

TABLE II. FREQUENT RULE ITEMS AND CANDIDATE RULE ITEMS

| 1st Pass | F1 | <({(A, e)}, 4), ((C, y), 3)>, <({(A, g)}, 5), ((C, y), 2)>, <({(A, g)}, 5), ((C, n), 3)>, <({(B, p)}, 3), ((C, y), 2)>, <({(B, q)}, 5), ((C, y), 3)>, <({(B, q)}, 5), ((C, n), 2)>, <({(B, w)}, 2), ((C, n), 2)> |
|---|---|---|
| 2nd Pass | C2 | <{(A, e), (B, p)}, (C, y)>, <{(A, e), (B, q)}, (C, y)>, <{(A, g), (B, p)}, (C, y)>, <{(A, g), (B, q)}, (C, y)>, <{(A, g), (B, q)}, (C, n)>, <{(A, g), (B, w)}, (C, n)> |
| | F2 | <({(A, e) , (B, p)}, 3), ((C, y), 2)>, <({(A, g) , (B, q)}, 3), ((C, y), 2)>, <({(A, g) , (B, q)}, 3), ((C, n), 1)>, <({(A, g) , (B, w)}, 2), ((C, n), 2)> |

TABLE III. CLASS ASSOCIATION RULES (CARs)

| CAR1 | (A, e) → (C,y), (A, g) → (C,n), (B, p) → (C,y), (B, q) → (C,y), (B, w) → (C,n) |
|---|---|
| CAR2 | {(A, e), (B, p)} → (C, y), {(A, g), (B, q)} → (C, y) {(A, g), (B, w)} → (C, n) |
| CARs | CAR1 U CAR2 |

TABLE IV. CLASS ASSOCIATION RULES AFTER PRUNING (PRCARS)

| prCAR1 | (A, e) → (C,y), (A, g) → (C,n), (B, p) → (C,y), (B, q) → (C,y), (B, w) → (C,n) |
|---|---|
| prCAR2 | {(A, g), (B, q)} → (C, y) |
| prCARs | prCAR1 U prCAR2 |

TABLE V. GENERATED RULES

| Rule | Support | Confidence |
|---|---|---|
| (1) A = e → y | 3/10 | 3/4 |
| (2) A = g → n | 3/10 | 3/5 |
| (3) B = p → y | 2/10 | 2/3 |
| (4) B = q → y | 3/10 | 3/5 |
| (5) B = w → n | 2/10 | 2/2 |
| (6) A = g, B = q → y | 2/10 | 2/3 |

## IV. PROPOSED MODELS

### A. The hardcoded model (CBA-ODM1)

In this model, the CBA is created starting from association rules and hardcoding the rest of the CBA algorithm in Oracle using Oracle PL/SQL. We call this model the Classification Based on Association built on Oracle Data Mining package (CBA-ODM1). The CBA-ODM1 is an association rules model that will be available in ODM package. Assuming that data is already cleaned and discretized, Figure 1 summarizes our implementation of this version into these steps :

1. Build association rules using ODM package by invoking Oracle ODM create a model function with a mining function called association.
2. Rank the rules based on confidence, support, and precedence of rule generation according to CBA criteria. Given two rules R1 and R2, R1 precedes R2 if:
    a. R1 confidence is higher than R2 confidence.
    b. R1 confidence equals R2 confidence, but R1 support is higher than R2 support.
    c. R1 confidence and support are equal, but R1 is generated before R2.
3. Build the classifiers and get the default class. We correlate each row items in the training dataset with each ranked rule row items, if we find that any attribute in the training data matches any of the ranked rules attributes we check the class value, if it is the same then we mark a classifier otherwise we loop till we find a classifier or all dataset is processed.
4. Evaluate the model.

Let TR denotes the Training Dataset, TS denotes the Testing Dataset, and T denotes a tuple in a dataset. Figure 2 shows the pseudo codes that illustrate the classifier builder of CBA-ODM1.

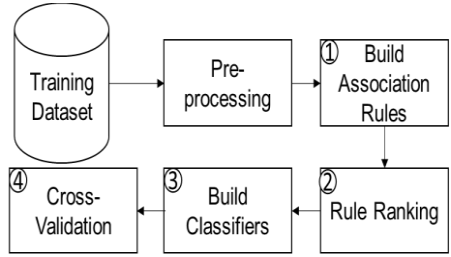

Fig. 1 Proposed CBA-ODM1 Model

```
1   ARM= ODM.Create_Model(mining_function=>Association);
2   ApprioriRules= ODM.Get_association_rules(ARM);
3   Rules_with_class = Filter(ARM on consequent = Target class);
4   Ranked_rules= Rank_rules(Rules_with_class);
5   For each T in TR do
6     For each Rule in Ranked_rules do
7       If T.Any_attribute_value = Rule.Attribute_value then
8         If T.class_value = Rule.Class_value then
9           If not exits classifier in classifers_list then
10             Insert rule.* into classifiers_list;
11           End if;
12           Exit;
13         End if;
14       End if;
15     End for;
16   Defualt Class = majority Class in TR
17  End for;
```

Fig. 2 The pseudo code of CBA-ODM1 Classifier Builder.

Line 1 creates association rules using the ODM data mining package which is equivalent to the Apriori algorithm in Oracle. Line 2 extracts the association rules from the model. Line 3 filters the list of association rules to have rules of the form X→C, where X is a list of items and C is a class value. Line 4 ranks rules according to the CBA ranking strategy; confidence, support and rule generation. In Line 5 we loop through all training data using ten-fold cross-validation. Line 6 loops through the rules generated in step 4. Lines 7-14 checks if a training record matches a rule then the rule is marked and added to the end of the classifiers.

To test this model, we have created a cross-validation procedure. The Pseudo code for this procedure is shown in Figure 3.

```
1   RandomRowsList =Random(TR);
2   For each row in RandomRowsList
3     PartionNumber = Mod (rownumber , nfolds);
4     PartitionedRandomRow= Concatenate (RandomRows , PartionNumber);
5   End for;
6   maxbucket = nfolds-1;
7   total_err=0;
8   for Iter in 0 to maxbucket loop
9     create_CBA_Model(TRi);
10    Error_rate= CBA_Model.get_error_rate;
11    total_err = total_err + Error_rate;
12  end for;
13  Average_error_rate= total_err/ nfolds;
```

Fig. 3 Pseudo code for cross-validation of CBA-ODM1.

Line 1 gets a list of the stratified training dataset. Line 2-5 assign a partition to each training data. Line 6 assigns the maximum bucket to a total number of folds minus 1. Line 7 initiates the total errors to 0. Line 8-13 calculate the average error for each created model.

### B. The integration model(CBA-ODM2)

We call this approach the simple or integration approach since it implements the CBA based on Apriori algorithm and Decision trees available in ODM package. We assume that data is already cleaned and discretized. We summarize our implementation of this version into these steps as shown in Figure 4:

1. Build association rules using the ODM package by invoking Oracle ODM create a model function with a mining function called association.
2. Generate the decision tree using the training data (9/10) using ten-fold cross-validation. We do not use any costing matrix for the decision tree we are assuming that all attributes have the same ratio of withering classified or not.
3. Convert the decision tree to a set of rules. This process involves processing complicated XML file because Oracle saves decision tree model details in XML.
4. Compare association rules with decision tree rules:
   a. If the decision tree is empty or it has one leaf, we neglect the tree and generate the classifier directly from ranked rules without pruning.
   b. If the decision tree has a match with attribute/value pairs of association rules, then choose the consequent of the higher confidence rule.
   c. Prune rules that do not have a match with any rule in the decision tree.
   d. Evaluate the model.

The pseudo code of the CBA-ODM2 is shown in Figure 5. Line 1 creates the association rules model. Line 2 gets the association rules. Line 3 creates a tree using the default tree setting table. In Line 3, the tree is converted to a set of rules. Line 4-8 is the combination step of rules. Finally, the default class is retrieved (Line 9).

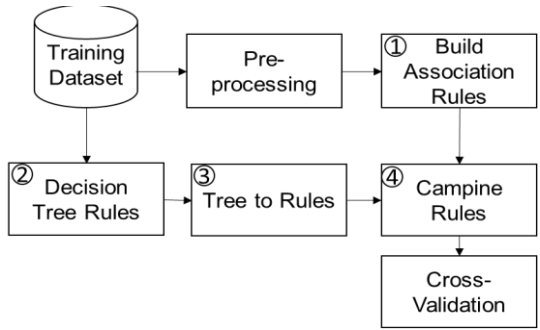

Fig. 4 Model for CBA_ODM2

```
1  ARM=ODM.Create_Model(mining_function=>Association);

2  ApprioriRules = ODM.Get_association_rules(ARM);

2  Tree_Model= ODM.Create_Model(mining_function=>classification,
                    Setting_table=> Tree_setting_Table);

3  DT_tree_rules =Convert2Rules(Tree_model);

4  If count(DT_tree_rules) >0 then

5      Select Rules_with_class ,
           case when dt.confidence > rr.confidence then
                dt.consequent else rr.consequent as consequent

6      From Apprio_rules rr, Decision_tree_rules DT

7      Where rr.attribute_name=Dt.attribute_name and

8          rr.attribute_value=dt.attribute_value ;

8  else Clasifier_lsit = Ranked_rules;

9  defualt class = majority class;
```

Fig. 5 Pseudo code for CBA-ODM2

## V. EXPERIMENTS AND EVALUATION

### A. Benchmark Dataset

The University of California, Irvine (UCI or UC Irvine), has contributed to the data mining community by its collection of databases known as "The UCI Machine Learning Repository" [33]. We choose 18 different datasets in terms of the number of instances, attributes, and the number of classes. The datasets are: Adult Income, Car Evaluation, Credit Default, Dermatology, Diabetic Retinopathy, E. coli, EEG, Haberman's Breast Cancer Survival, Ionosphere Radar Returns, Mice Protein Expression, Nursery Admittance, Seed Classification, Seismic, Soybean, Teaching Assistant Evaluation, Tic Tac Toe Endgame, Website Phishing, and Wholesale Customer Region.

### B. Selected Methods and Scenarios

We run the comparison against the following methods- Support Vector Machine, Naive Bayes, Decision Trees, Random Forests. The standard implementations in the *sklearn* library[34] were adopted. The CBA was run using the python implementation pyARC of [35]. By using 10-fold cross-validation after shuffling, we use nine datasets for training and one for testing using several scenarios to justify proper model evaluation on different parameters' values.

1. minimum support of 35% and minimum confidence to 50%.
2. minimum support of 15% and minimum confidence to 50%.
3. minimum support of 10% and minimum confidence to 50%.
4. minimum support of 5% and minimum confidence to 50%.

### C. Experiments on Selected Methods

The previous methods were applied to the selected UCI datasets. Table VI shows the average accuracy of the four chosen methods. Since the Decision tree algorithm was getting the highest accuracy, we compare it with the CBA and its proposed variants as shown in Table VII.

TABLE VI. THE AVERAGE ACCURACY OF SELECTED METHODS

|  | Average Accuracy |
|---|---|
| **Support Vector** | 0.76 |
| **Naive Bayes** | 0.68 |
| **Decision Tree** | **0.77** |
| **Random Forests** | 0.73 |

TABLE VII. COMPARISON OF CBA VARIANTS

|  | CBA | CBA-ODM1 | CBA-ODM2 |
|---|---|---|---|
| Scenario 1 | 0.83 | 0.79 | 0.84 |
| Scenario 2 | 0.78 | 0.74 | 0.78 |
| Scenario 3 | 0.77 | 0.75 | 0.79 |
| Scenario 4 | 0.79 | 0.78 | 0.78 |
| **Average** | 0.79 | 0.77 | **0.80** |

As Table VII shows, the performance of CBA-ODM2 is 1% higher than the original CBA algorithm due to the usage of the integration methods with decision trees and oracle association rules. However, the CBA-ODM1 was 2% lower than the original CBA method due to the adopted pruning technique. As a result, the CBA and its variants outperform the decision tree method with 2-3%.

We investigate the accuracy of our model relative to the number of attributes. We take the average accuracy of the last two scenarios, and then we categorize attributes in 5 groups (4-10 attributes, 11-20 Attributes, 30-50, more than 50 attributes). The CBA-ODM2 outperforms the CBA for datasets that have 6,8,21,34 attributes with increase in accuracy of (1.5%,2.2%,1.1%,0.9%)respectively. The conclusion is that as the number of attributes increase, there will be many permutations were some of them is not explainable. The performance of Adult Income dataset was the worst for all compared methods.

The relationship between the dataset size and accuracy is also investigated. We group datasets into three groups: datasets with less than 1000 instances, tables with instances more than 1000 and less than 5000 and tables with more than 5000 instances. The reason for grouping is trying to find a relation between groups rather than a single data set. Results showed that as the number of instances increases, the proposed classifier(CBA-ODM1) becomes more accurate because the classifier is being built on a more representable data.

We also considered in another test the number of classes for each instance and us group datasets into 2,3,4,5,8 classes. Experiments showed that the fewer classes, the more likely the data would be predicted correctly. However naive Bayes is achieving better with more classes. Our tests showed that 51% of class labels are matched between the Apriori and Decision tree (CBA-ODM2).

We noticed that the implementation of CBA-ODM2 outperforms CBA-ODM1 due to enhancing performance with the integration of decision trees algorithm. Moreover, we get results from CBA-ODM2 model faster than CBA-ODM1.

Since the proposed CBA-ODM2 has three disjoint parts, we can run this model in parallel, and so large datasets should not be a big issue.

The implication of this work is practical and theoretical. Practically adding a new algorithm to a well-known and scalable database will increase the productivity of enterprises especially those working on OLAP. Theoretically, the model creates a new set of classifiers by integrating two models, the decision tree and the Apriori method.

This article has a set of limitations. The model has been applied on 18 datasets; therefore, researchers must interpret results accordingly before generalization. Moreover, the model is not yet implemented physically in Oracle source code; therefore, we plan to place the code as an addon on Oracle Data miner.

## VI. CONCLUSION

From this work we have found that our CBA models resample a commercial and state of the art CBA algorithms. We have prepared a standalone Oracle package that can be used in Oracle package easily by a single command. We have integrated decision trees with Apriori in Oracle and compared results with a set of classification methods. Results showed that the proposed algorithms outperform chosen methods with an increase of 1% in accuracy. The generated rules of the CBA-ODM1 are similar to those produced by original CBA; however, there is some difference which is in the range of 1-2% as a result of improvements. The ODM-CBA2 rules are leaned to those produced by the decision tree which results in improvements in the proposed model.